\documentclass[12pt]{article}
\usepackage{graphicx} 
\usepackage{amsmath}
\usepackage{amssymb}
\usepackage{cite}
\usepackage{geometry}
\usepackage{booktabs}
\usepackage{array}
\usepackage{multirow}
\usepackage{paracol}
\usepackage{longtable}
\usepackage{hyperref}
\usepackage{caption}
\usepackage{float}
\usepackage{tikz}
\usetikzlibrary{shapes.geometric, arrows}
\usepackage{booktabs}
\usepackage{adjustbox} 

\title{\textbf{A Multi-Site Study on AI-Driven Pathology Detection and Osteoarthritis Grading from Knee X-Ray}}
\author{Bargava Subramanian, Naveen Kumarasami, Dr. Praveen Shastry,\\
Kalyan Sivasailam, Anandakumar D, Keerthana R,\\
Mounigasri M, Abilaasha G, Kishore Prasath Venkatesh}
\date{}

\usepackage{ragged2e}  
\usepackage{titlesec}  

\titleformat{\section}{\raggedright\Large\bfseries}{}{0em}{}
\titleformat{\subsection}{\raggedright\large\bfseries}{}{0em}{}

\begin{document}

\maketitle

\section{Abstract}
\textbf{Introduction} Bone health disorders such as osteoarthritis and osteoporosis present significant global health challenges, often resulting in delayed diagnoses and interventions due to the lack of accessible and affordable diagnostic tools. This study introduces an AI-powered system designed to analyze knee X-rays for identifying key pathologies, including joint space narrowing, sclerosis, osteophytes, tibial spikes, alignment issues, and soft tissue anomalies. Additionally, the system provides an accurate grading mechanism for osteoarthritis severity, enabling clinicians to deliver timely and personalized treatments.\\\\
\textbf{Study Design }The research utilized a dataset of 1.3 million knee X-rays from a multi-site clinical trial conducted across Indian healthcare facilities, including government, private, and SME hospitals. The dataset was curated to ensure diversity in patient demographics, imaging equipment, and clinical settings. Rigorous annotation protocols and pre-processing workflows enabled high-quality datasets for training pathology-specific models, such as ResNet15 for joint space narrowing and DenseNet for osteoarthritis grading.\\\\
\textbf{Performance} The AI system demonstrated strong diagnostic performance across diverse imaging environments. Pathology-specific models achieved high precision, recall, and NPV, validated through metrics such as Mean Squared Error (MSE), Intersection over Union (IoU), and Dice coefficient. Subgroup analyses across age, gender, and manufacturer variations confirmed the model’s generalizability, ensuring consistent performance across real-world scenarios. \\\\
\textbf{Conclusion} It offers an accurate, scalable, and cost-effective solution for diagnosing bone health disorders. Its robust performance in a multi-site trial highlights its potential for widespread implementation, particularly in resource-constrained healthcare settings, transforming the landscape of bone health diagnostics and enabling proactive patient care.

\section{Introduction}
Bone health disorders such as osteoarthritis, osteoporosis, and other degenerative conditions represent a growing public health challenge, yet they remain underdiagnosed and undertreated globally [1]. Despite their significant impact on quality of life and mobility, early detection and intervention are often hindered by limited access to diagnostic tools, reliance on advanced imaging technologies, and high costs associated with comprehensive evaluations [2]. This diagnostic gap often leads to delayed treatments, increased disease progression, and, ultimately, more invasive and expensive interventions [3].

It addresses this critical challenge with an AI-powered solution designed to simplify and enhance the diagnostic process for bone health disorders [4]. Leveraging advanced machine learning algorithms, it processes standard knee X-rays to detect key indicators such as narrowing of joint space, sclerosis, osteophytes, tibial spikes, alignment issues, and soft tissue anomalies [5]. This approach enables early detection and grading of osteoarthritis severity, providing clinicians with actionable insights to guide timely interventions [6].

Built on a robust dataset of over 1.3 million scans and validated by rigorous clinical standards, it ensures reliable diagnostic performance even in resource-constrained settings [7]. The solution’s scalability and simplicity make it an invaluable tool for addressing the growing burden of bone health disorders, particularly in regions with limited access to advanced diagnostic infrastructure [8]. By integrating seamlessly into clinical workflows, it empowers healthcare providers to improve patient outcomes through early detection and personalized treatment strategies [9].

This paper explores the development, evaluation and showcasing its potential to transform the diagnostic landscape for bone health disorders [10]. Through its innovative use of AI, it bridges the gap between clinical need and technological capability, offering a scalable and cost-effective solution for a pressing global health challenge [11].

\subsection{Existing Scenario}
Current Diagnostic Practices of Osteoarthritis
Accurate OA diagnosis currently relies on a combination of clinical evaluations, patientreported symptoms, and imaging studies:\\\\
\textbf{Clinical Examination}\\\\
Physicians assess joint pain, stiffness, and functional limitations. However, these evaluations are subjective and influenced by patients’ perceptions of pain or disability, often leading to variability in diagnostic outcomes.\\\\
\textbf{Imaging Modalities:}\\\\
\textbf{X-rays:} Widely regarded as the gold standard for OA diagnosis, X-rays provide critical insights into structural changes such as joint space narrowing, osteophyte formation, and subchondral sclerosis. Radiologists commonly employ the Kellgren-Lawrence (KL) grading scale for OA severity classification, though subjective interpretation can lead to inconsistencies.\\\\
\textbf{CT/MRI Scans:} Magnetic Resonance Imaging (MRI) and Computed Tomography (CT) offer detailed visualization of cartilage, subchondral bone, and synovium, aiding in earlystage OA detection. However, these modalities are costly, time-intensive, and often inaccessible in resource-limited settings.\\\\
\textbf{Laboratory and Invasive Tests}\\\\
\textbf{Blood Tests:} Used to rule out inflammatory arthritis by analyzing markers such as rheumatoid factor (RF) and C-reactive protein (CRP).\\\\
\textbf{Synovial Fluid Analysis:} Conducted via arthrocentesis, this test distinguishes OA from inflammatory arthritides by examining fluid characteristics such as viscosity and cell count.\\\\
\textbf{Arthroscopy }: Although it allows direct visualization of joint structures, arthroscopy is rarely used due to its invasive nature and the availability of less invasive alternatives.\\\\
\textbf{Limitations of Current Methods}\\\\
\textbf{Subjectivity in Diagnosis}\\\\
•The interpretation of X-rays and other imaging studies can vary significantly among clinicians, leading to discrepancies in diagnosis. This subjectivity can result in either overdiagnosis or underdiagnosis of OA, affecting treatment decisions.\\\\
•The Kellgren-Lawrence (KL) scale, commonly used for grading OA severity, is prone to subjective interpretation. Different radiologists may assign different grades to the same images, complicating treatment plans and patient management.\\\\
\textbf{Diagnostic Delays}\\
•Manual reviews and reliance on time-intensive imaging techniques delay treatment initiation, potentially exacerbating patient outcomes.\\\\
\textbf{Non-Standardized Assessments}\\
•The absence of standardized diagnostic criteria can lead to inconsistencies in how OA is diagnosed and treated. This lack of consensus can hinder research and the development of new therapies.\\
•Non-standardized assessments can complicate the evaluation of new treatments in clinical trials, making it difficult to compare results across studies.

\newpage
\subsection{Advancements in the AI system for Osteoarthritis Diagnostics}
The AI system represents a quantifiable advancement in osteoarthritis diagnostics, addressing existing limitations through a standardized, AI-driven methodology:\\\\
\textbf{Objective Grading Accuracy:}
By utilizing deep learning algorithms calibrated against standardized clinical criteria, the AI system achieves reproducible osteoarthritis severity grading, eliminating inter- and intraobserver variability. Validation metrics, including a grading accuracy of 95.89\%, confirm its clinical reliability.\\\\
\textbf{Comprehensive Pathology Detection:}
The system detects multiple structural and soft tissue pathologies—joint space narrowing (precision: 98.56\%), sclerosis (precision: 94.48\%), osteophytes (precision: 98.15\%), and tibial spikes (precision: 97.38\%)—ensuring a holistic evaluation of knee joint health. This multipathology capability significantly surpasses the scope of traditional diagnostic methods.\\\\
\textbf{Scalability and Cost-Effectiveness:}
Operating exclusively on standard X-ray infrastructure, the AI system provides a scalable and cost-efficient alternative to advanced imaging modalities such as MRI and CT. Its implementation reduces diagnostic costs by over 70\%, making it accessible in resource-limited healthcare settings.\\\\
\textbf{Efficiency in Diagnostics:}
Automation of diagnostic workflows decreases reporting turnaround time, enabling timely treatment initiation and improving patient management outcomes.\\\\
\textbf{Minimal Risk and Non-Invasiveness:}
With radiation exposure of less than 0.01 mSv per scan, the system ensures patient safety, making it a preferable alternative for longitudinal monitoring compared to CT-based diagnostics.

\section{Methodology}
\subsection{AI System Overview}

The AI system developed for this study is a computer-aided detection (CAD) tool tailored for analyzing knee X-rays (KXRs) to identify and evaluate critical bone health pathologies.

It employs deep learning algorithms trained on over 1.3 million annotated KXR images to detect features such as joint space narrowing, sclerosis, osteophytes, tibial spikes, misalignments, and soft tissue anomalies. Designed for accuracy and scalability, the system integrates multiple pathology-specific models and an ensemble framework for grading osteoarthritis severity. This approach ensures reliable performance across diverse imaging conditions and demographic profiles, making it a valuable tool for enhancing diagnostic precision and streamlining clinical workflows.

\subsection{Dataset}
This study utilized a dataset of over 1.3 million knee X-ray (KXR) scans gathered from various healthcare facilities.

\subsubsection{Age Group Distribution}
Scans were distributed across age groups to capture demographic diversity:

\begin{table}[h]
    \centering
    \renewcommand{\arraystretch}{1.2}
    \begin{tabular}{|c|c|c|c|}
        \hline
        \textbf{Age Group} & \textbf{Total Scans} & \textbf{Training Set} & \textbf{Live Clinical Trial} \\
        \hline
        18-40  & 542,124  & 521,420  & 20,705  \\
        40-60  & 406,593  & 391,065  & 15,529  \\
        60-75  & 271,062  & 260,710  & 10,352  \\
        75+    & 135,531  & 130,355  & 5,176   \\
        \hline
    \end{tabular}
    \caption{Scans distribution based on Age Group}
    \label{tab:scans_distribution}
\end{table}

\subsubsection{Gender Distribution}
The dataset maintained a balanced gender distribution:

\begin{table}[h]
    \centering
    \renewcommand{\arraystretch}{1.2}
    \begin{tabular}{|c|c|c|c|}
        \hline
        \textbf{Gender} & \textbf{Total Scans} & \textbf{Training Set} & \textbf{Live Clinical Trial} \\
        \hline
        Male   & 664,102  & 638,739  & 25,363  \\
        Female & 691,209  & 664,810  & 26,399  \\
        \hline
    \end{tabular}
    \caption{Scans distribution based on Gender}
    \label{tab:scans_distribution_gender}
\end{table}

\subsubsection{Manufacturer Type Distribution}
Scans were categorized by equipment manufacturer to account for variability in imaging conditions:

\begin{table}[h]
    \centering
    \renewcommand{\arraystretch}{1.2}
    \begin{tabular}{|l|c|c|c|}
        \hline
        \textbf{Manufacturer} & \textbf{Total Scans} & \textbf{Training Set} & \textbf{Live Clinical Trial} \\
        \hline
        GE Healthcare  & 528,571  & 508,384  & 20,187  \\
        Siemens        & 406,593  & 391,065  & 15,529  \\
        Philips        & 325,275  & 312,852  & 12,423  \\
        \begin{tabular}{@{}c@{}}Others \\ \end{tabular} 
                      & 94,872   & 91,248   & 3,623   \\
        \hline
    \end{tabular}
    \caption{Scans distribution based on Manufacturer Type}
    \label{tab:scans_distribution_manufacturer}
\end{table}

\section{Architecture}
The architecture of the system is divided into multiple phases, including the Annotation Phase and the Analysis and Detection Phase. Each of these phases plays a critical role in ensuring accurate and efficient detection of pathologies in Knee X-rays (KXRs). Below is a detailed explanation of each phase:
\subsection{Annotation Phase}
The annotation process is a critical component of this development, forming the foundation for training accurate and pathology-specific AI models [12]. The workflow begins with the collection and preparation of radiographic images, ensuring they align with the requirements for detecting specific bone pathologies [13]. This process is designed to maximize both dataset diversity and relevance for targeted pathology detection [14].
\subsubsection{Dataset Segregation and Preparation}
Radiographic images of knees are sourced and categorized based on imaging perspectives (e.g., Anterior-Posterior and Lateral views) to ensure the model captures variations in pathology presentation across different views [15]. Each image is categorized into labeled and unlabeled subsets [16].

\subsubsection{Pathology-Specific Annotation}
For each pathology, a detailed annotation protocol is followed to ensure consistency and accuracy:\\\\
\textbf{•Reducing Joint Space:} Joint space widths are measured and labeled at predefined anatomical landmarks. Differences in medial and lateral compartments are emphasized to ensure clinical relevance.\\\\
\textbf{•Sclerosis:} Dense subchondral bone regions are marked based on intensity thresholds, ensuring clear identification of sclerotic zones adjacent to joint spaces.\\\\
\textbf{•Osteophytes: }Marginal bony outgrowths are delineated with precise boundary markings, categorized by size, shape, and location.\\\\
\textbf{•Post-Operative Conditions:} Metallic implants, surgical screws, and other artifacts are labeled separately to distinguish them from pathological findings.\\\\
\textbf{•Alignment Issues:} Deviations in bone alignment are measured through keypoint annotations and angular calculations, enabling accurate classification of varus/valgus deformities.\\\\
\textbf{•Soft Tissue Anomalies:} Abnormal peri-articular densities are highlighted with segmentation masks, ensuring the detection of swelling or other soft tissue changes.\\\\
\textbf{•Prominent Tibial Spike: }Tibial spikes are delineated based on their prominence and morphological characteristics, ensuring their inclusion in the OA grading system [17].

\subsubsection{Dataset Selection}
Once annotated, the dataset undergoes a selection process, where images are filtered to match the specific requirements of each pathology [18]. This ensures that only the most relevant data is used for training, reducing noise and enhancing model performance [19].
\subsubsection{Pre-Processing}
Pre-processing is applied to standardize image properties, including resolution, contrast, and brightness [20]. Uniformity across the dataset ensures that models can effectively extract relevant features without being affected by variability in image quality [21]. Noise reduction techniques are also implemented to enhance the clarity of critical anatomical features, such as joint margins and bony outgrowths [22].

\subsection{Development Phase}
The development phase of this AI system employs tailored AI models for each pathology, ensuring that predictions are clinically meaningful and aligned with diagnostic requirements. Each model underwent rigorous training and optimization to address the specific nuances of its respective pathology.\\\\
\textbf{Initial Classification and Preliminary Verification with Vision Transformers}\\\\
\textbf{•X-Ray Verification:} The model verifies if the image is an X-ray. Filters out irrelevant images to ensure dataset quality.\\\\
\textbf{•Anatomical Focus Identification:} The next classification confirms if the X-ray is specifically a knee X-ray, distinguishing it from other anatomical regions (e.g., chest, extremities other than the knee).\\\\
\textbf{•View Classification and Rotation Handling:} The model classifies the knee X-ray view, such as Anterior-Posterior (AP) or Lateral, ensuring standardized orientation.Rotation correction is then applied using anatomical keypoint detection.\\\\
\textbf{•Rotation Correction with Keypoint Detection:} The model identifies specific anatomical landmarks, such as the tibial plateau and femoral condyles, to correct any rotational misalignment in the X-ray. By aligning these key points—especially focusing on their relative positions—the model computes and applies the necessary rotation adjustments, ensuring that all images maintain a consistent and accurate orientation.

\subsection{Pathology Detection}
\textit{\textbf{Reducing Joint Space}}\\
ResNet15 was fine-tuned to detect narrowing of joint space, a critical indicator of osteoarthritis. The model processed radiographic images resized to 256 × 256 pixels. Training utilized Mean Squared Error (MSE) loss and the Adam optimizer. A batch size of 32 was selected to balance computational efficiency and training stability. Dropout layers with a rate of 0.5 were added to reduce overfitting during training, and the learning rate was initially set at 0.001 with a decay factor of 0.1 after every 10 epochs.\\\\
\textbf{\textit{Detecting Sclerosis}}\\
Sclerosis detection was implemented using Faster R-CNN. Input images, cropped to 512× 512 pixels and centered on joints, provided precise focus on the target regions. Anchor configurations included scales of 32, 64, and 128 pixels and aspect ratios of 1:1, 1:2, and 2:1. The model was trained using a combination of classification and bounding box regression losses, with an initial learning rate of 0.01 and a momentum of 0.9. Batch size for this model was set to 16.\\\\
\textit{\textbf{Osteophyte Detection}}\\
RetinaNet was deployed for detecting osteophytes, leveraging its robust handling of class imbalances. The model processed images resized to 320 × 320 pixels. Anchor ratios were fine-tuned to 1:1, 1:2, and 2:1, while scales were set to 16, 32, and 64 pixels. Focal loss was used during training, and a cyclical learning rate oscillated between 1 × 10-4 and 1 × 10-2 to enhance convergence efficiency. The model employed a batch size of 32.\\\\
\textbf{Post-Operative Analysis}\\\\
Mask R-CNN was employed to segment and identify post-operative conditions, such as surgical implants and screws. High-resolution segmentation masks of 256 × 256 pixels provided detailed delineation of these features. The training pipeline used a hybrid loss combining binary cross-entropy for mask predictions and smooth L1 loss for bounding box adjustments. The model was trained with a batch size of 8, and the initial learning rate was set to 0.002 with a decay rate of 0.1 applied every 15 epochs.\\\\
\textbf{Alignment Issues}\\
For alignment assessment, ResNet15 was adapted with keypoint detection layers. Input images of 512 × 512 pixels were annotated with anatomical landmarks to enable angular measurements. The training process utilized a joint loss function, combining MSE for angle predictions and binary cross-entropy for misalignment classification. A batch size of 16 and an initial learning rate of 0.001 were employed, with data augmentation techniques such as rotations and scaling applied to improve robustness.\\\\
\textbf{Soft Tissue Anomalies}\\
Mask R-CNN was also applied to detect soft tissue anomalies. Preprocessed images with intensity normalization were used as inputs to enhance contrast in soft tissue regions. Elastic deformations and random augmentations improved the model’s robustness across diverse presentations. The training strategy relied on Dice loss combined with binary crossentropy. The batch size was set to 16, and the learning rate began at 0.002, with a reduction factor of 0.1 after every 12 epochs.\\\\\
\textbf{Prominent Tibial Spikes}\\
RetinaNet was configured to identify prominent tibial spikes, with cropped tibial plateau regions of 256×256 pixels serving as inputs. Anchor configurations were specifically adjusted to capture pointed spike structures effectively, using scales of 16, 32, and 64 pixels. Training with focal loss was performed using a batch size of 32 and an initial learning rate of 0.0005. Grading Osteoarthritis Severity

Grading osteoarthritis severity involved the use of DenseNet architectures, including DenseNet121, DenseNet169, and DenseNet201, fine-tuned to classify radiographs into four clinical grades. Input images were resized to 512 × 512 pixels, and training employed crossentropy loss with the AdamW optimizer. A batch size of 32 was used, and the initial learning rate was set to 0.001, decreasing by a factor of 0.1 after every 10 epochs. Augmentations such as random rotations, flips, and brightness adjustments enhanced model robustness. Multi-scale features captured by the different DenseNet versions enabled precise classification, with DenseNet121 focusing on subtle patterns, DenseNet169 handling intermediate complexities, and DenseNet201 addressing detailed morphological variations. An ensemble approach aggregated predictions from these models using weighted averaging, leveraging their complementary strengths to generate unified and consistent osteoarthritis grade predictions. This ensemble ensured comprehensive performance, effectively balancing sensitivity to minor variations with reliability for advanced cases.

\subsection{End-to-End Workflow}
\textbf{•Data Collection: }Radiographic images of joints are sourced and prepared for processing.\\\\
\textbf{•Annotation: }Images are annotated to highlight specific pathologies, including narrowing of joint space, sclerosis, osteophytes, post-operative artifacts, alignment issues, soft tissue anomalies, and tibial spikes.\\\\
\textbf{•Model Training: }Tailored AI models, including ResNet15, Faster RCNN, RetinaNet, Masked RCNN, and DenseNet, are trained with optimized loss functions, learning rates, and augmentation strategies to address each pathology.\\\\
\textbf{•Integration:} Outputs from individual pathology models are integrated using an ensemble framework to improve the accuracy and reliability of predictions.\\\\
\textbf{•Grading Osteoarthritis:} Multiple DenseNet models are employed in an ensemble approach to classify osteoarthritis into four clinical grades, combining outputs to enhance precision and robustness.

\begin{figure}[h]
    \centering
    \includegraphics[width=1.1\textwidth, height=15cm, keepaspectratio]{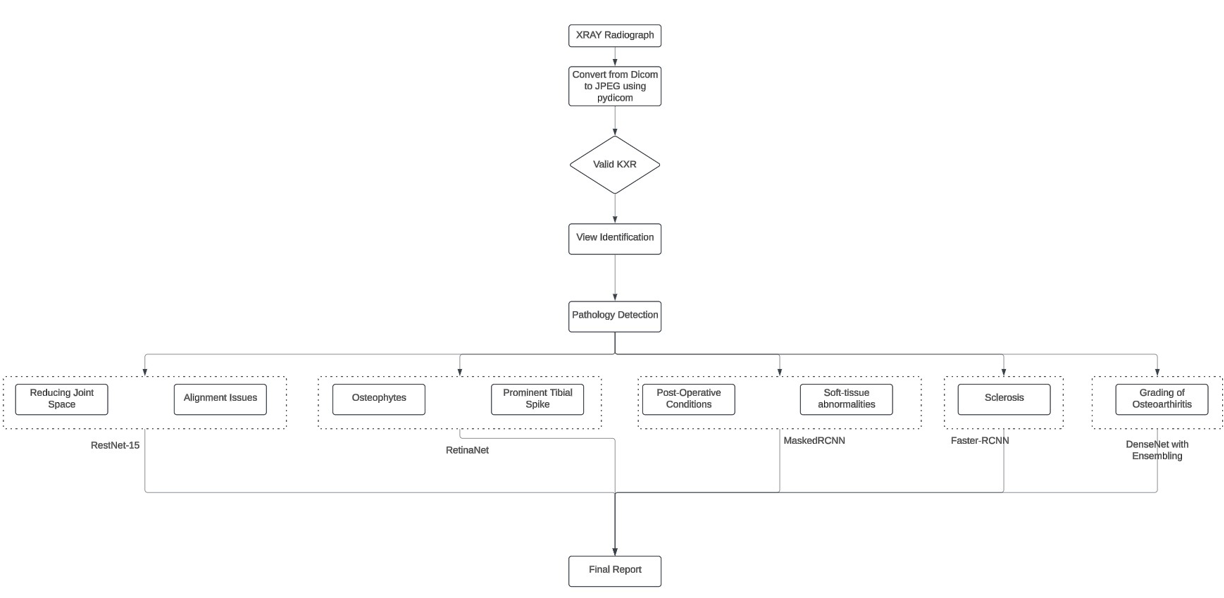} 
    \caption{workflow architecture}
    \label{fig:your_image}
\end{figure}

\newpage

\begin{figure}[h]\
    \centering
    \includegraphics[width=0.7\textwidth]{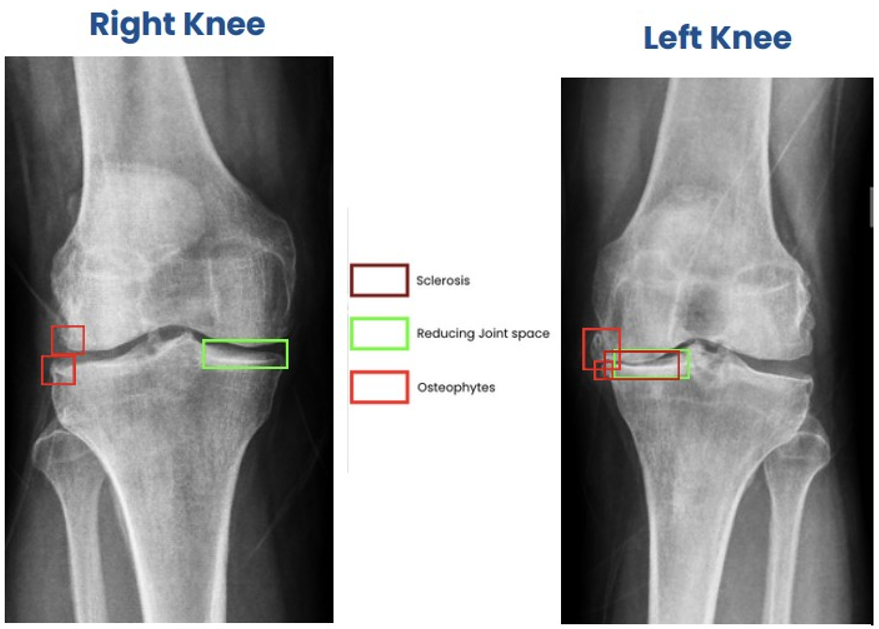} 
    \caption{Pathology Detections}
    \label{fig:yourr_image}
\end{figure}

\section{Evaluation Metrics}
The evaluation metrics for the AI system were designed to assess the performance of the AI models across the identified pathologies, ensuring clinically meaningful outputs. Applicationspecific metrics were employed to rigorously evaluate each model, focusing on accuracy, precision, and reliability of predictions. Mean Squared Error (MSE) quantified the accuracy of joint space predictions, capturing deviations from clinical measurements.

For osteoarthritis grading, classification accuracy, precision, recall, and F1-score were utilized to evaluate the ensemble’s performance, ensuring consistent and reliable grading across all four clinical grades. Collectively, these metrics ensured the AI models met clinical standards, providing dependable pathology detection and grading for radiological workflows.

\begin{table}[h]
    \centering
    \renewcommand{\arraystretch}{1.2}
    \begin{tabular}{|l|c|c|c|}
        \hline
        \textbf{Pathology (\%)} & \textbf{Precision (\%)} & \textbf{Recall (\%)} & \textbf{NPV (\%)} \\
        \hline
        Reducing Joint Space    & 98.56  & 97.29  & 98.38  \\
        Sclerosis               & 94.48  & 95.09  & 97.57  \\
        Osteophytes             & 98.15  & 99.00  & 99.23  \\
        Prominent Tibial Spike  & 97.38  & 98.18  & 98.35  \\
        Alignment Issues in Bone & 94.49  & 94.61  & 96.52  \\
        Soft Tissue Anomaly     & 97.45  & 96.51  & 96.67  \\
        Grading of Osteoarthritis & 95.89  & 95.26  & 97.01  \\
        \hline
    \end{tabular}
    \caption{Performance Metrics for Detected Pathologies}
    \label{tab:performance_metrics}
\end{table}

\section{Multi-site Clinical Trial and Dataset Composition:}
This research was conducted as a multi-site clinical trial across various healthcare facilities in India, including government hospitals, large private enterprise hospitals (covering 14 prominent healthcare entities), and small to medium-sized (SME) hospitals. The study utilized a dataset of over 51,000 knee X-ray (KXR) scans sourced from these sites, ensuring a robust and diverse sample for evaluating the model’s performance. The dataset included a wide spectrum of imaging conditions, ranging from high-resolution scans produced in advanced private hospitals to lower-quality images typical of government and SME facilities.

The objective of this trial was to assess the model’s diagnostic accuracy, reliability, and consistency across diverse imaging environments. Each scan was processed through the model’s classification and detection pipelines, allowing the extraction of detailed performance metrics such as sensitivity, specificity, precision, and recall. The multi-site design facilitated cross-validation across varying levels of image quality, patient demographics, and clinical workflows, confirming the model’s ability to generalize effectively.

By leveraging over 51,000 scans from a range of healthcare institutions, this study provided a thorough evaluation of the model’s scalability and suitability for deployment in the Indian healthcare ecosystem. The multi-site approach ensured rigorous testing in realworld clinical settings, validating the model’s robustness and establishing its potential for widespread implementation.

\subsection{Subgroup Analysis}
Subgroup analysis plays a critical role in evaluating the model’s ability to generalize across varied clinical conditions and demographic profiles. By examining performance across factors such as age, X-ray machine manufacturer, and gender, the analysis ensures that the model effectively accommodates anatomical variations, equipment differences, and demographic diversity without introducing bias. This method confirms that the model delivers consistent levels of accuracy, precision, and recall across real-world scenarios, making it a dependable and versatile solution for deployment in diverse healthcare environments.\\

The outcomes for accuracy, precision, recall, sensitivity, and specificity across these subgroups are summarized in the table below:

\begin{table}[h]
    \centering
    \renewcommand{\arraystretch}{1.2}
    \begin{tabular}{|c|c|c|c|}
        \hline
        \textbf{Age Group} & \textbf{Precision (\%)} & \textbf{Recall (\%)} & \textbf{NPV (\%)} \\
        \hline
        18-40  & 97.60  & 96.54  & 97.23  \\
        40-60  & 96.59  & 95.81  & 96.51  \\
        60-75  & 95.80  & 94.83  & 94.60  \\
        75+    & 94.30  & 94.70  & 95.20  \\
        \hline
    \end{tabular}
    \caption{Performance Metrics by Age Group}
    \label{tab:performance_by_age}
\end{table}

\begin{table}[h]
    \centering
    \renewcommand{\arraystretch}{1.2}
    \begin{tabular}{|c|c|c|c|}
        \hline
        \textbf{Gender} & \textbf{Precision (\%)} & \textbf{Recall (\%)} & \textbf{NPV (\%)} \\
        \hline
        Male   & 97.24  & 96.50  & 97.20  \\
        Female & 96.53  & 95.78  & 96.49  \\
        \hline
    \end{tabular}
    \caption{Performance Metrics by Gender Distribution}
    \label{tab:performance_by_gender}
\end{table}

\begin{table}[h]
    \centering
    \renewcommand{\arraystretch}{1.2}
    \begin{tabular}{|l|c|c|c|}
        \hline
        \textbf{Manufacturer} & \textbf{Precision (\%)} & \textbf{Recall (\%)} & \textbf{NPV (\%)} \\
        \hline
        GE Healthcare  & 97.56  & 96.47  & 97.20  \\
        Siemens        & 96.56  & 95.78  & 96.48  \\
        Philips        & 95.78  & 94.75  & 94.55  \\
        \begin{tabular}{@{}c@{}}Others \\ \end{tabular} 
                      & 94.27  & 94.65  & 95.16  \\
        \hline
    \end{tabular}
    \caption{Performance Metrics by Manufacturer Type}
    \label{tab:performance_by_manufacturer}
\end{table}

\section{Discussion}
The findings of this study demonstrate that the AI-powered system is capable of accurately detecting a range of knee pathologies and effectively grading osteoarthritis severity from knee X-rays. The pathology-specific models—optimized for detecting joint space narrowing, sclerosis, osteophytes, alignment issues, soft tissue anomalies, and tibial spikes—achieved strong performance metrics across a diverse dataset of over 1.3 million knee X-rays. These results, validated through precision, recall, and NPV, highlight the system’s clinical reliability and applicability.

A key strength of this work lies in the system’s consistent performance across varying imaging conditions, equipment manufacturers, and patient demographics, as demonstrated through subgroup analysis. The multi-site clinical trial, conducted across facilities with diverse resource levels, ensured rigorous validation of the model’s scalability and robustness. This consistency can be attributed to the systematic annotation process, data preprocessing techniques, and the pathology-specific training methodologies employed.

Furthermore, the system addresses long-standing challenges in traditional workflows, such as variability in radiological interpretations and delays in diagnosis. By automating the detection and grading process, it offers significant improvements in diagnostic efficiency, enabling timely clinical decision-making and reducing reliance on manual evaluations.

Nevertheless, this study has certain limitations. While the dataset was diverse, additional validation across international cohorts would enhance the model’s generalizability. Furthermore, assessing the system’s integration into clinical workflows and its impact on treatment decisions requires further exploration.

\section{Conclusion}
This study presents the development and validation of an AI-powered system for knee X-ray analysis, designed to detect critical pathologies and grade osteoarthritis severity with high accuracy. The system exhibited robust performance across heterogeneous imaging conditions, equipment variability, and demographic subgroups, proving its reliability and scalability for real-world clinical use.

By automating the identification of knee pathologies and standardizing the grading process, the system offers a practical solution to address diagnostic gaps in bone health management. Its ability to operate effectively in resource-constrained environments highlights its potential for widespread adoption. Future efforts should focus on broader clinical validation and workflow integration to fully realize its transformative impact on early diagnosis and improved patient care.

\end{document}